\documentclass[a4paper]{jpconf}
\usepackage{graphicx,amsmath,alltt,axodraw2,pifont}

\DeclareSymbolFont{ttoperators}{OT1}{cmtt}{m}{n}
\newcommand\xCode[1]{{%
  \mathcode`\"="0\the\symttoperators22%
  \mathchardef\$="4\the\symttoperators24%
  \mathcode`\(="4\the\symttoperators28%
  \mathcode`\)="5\the\symttoperators29%
  \mathcode`\/="0\the\symttoperators2F%
  \mathcode`\[="4\the\symttoperators5B%
  \mathcode`\]="5\the\symttoperators5D%
  \mathchardef\{="4\the\symttoperators7B%
  \mathchardef\}="5\the\symttoperators7D%
  \ensuremath{\mathtt{#1}}}}
\newcommand\Code[1]{\texttt{#1}}
\newcommand\Var[1]{{\it #1}}
\newcommand\eg{e.g.\ }
\newcommand\ie{i.e.\ }

\newcommand\MSbar{\overline{\text{MS}}}
\newcommand\DRbar{\overline{\text{DR}}}
\newcommand\uscore{\symbol{95}}
\newcommand\lbrac{\symbol{123}}
\newcommand\rbrac{\symbol{125}}
\newcommand\Brac[1]{\lbrac#1\rbrac}
\newcommand\Vi{\Var{i}}
\newcommand\Vj{\Var{j}}

\begin{document}

\title{FormCalc 9 and Extensions}

\author{T.~Hahn$^1$, S.~Pa\ss ehr$^2$, C.~Schappacher$^3$}

\address{%
${}^1$ Max-Planck-Institut f\"ur Physik,
       F\"ohringer Ring 6,
       D--80805 Munich, Germany, \\
${}^2$ DESY,
       Notkestr. 85,
       D--22607 Hamburg, Germany, \\
${}^3$ Institut f\"ur Theoretische Physik, KIT,
       D--76128 Karlsruhe, Germany
       (former address).}

\ead{hahn@mpp.mpg.de, sebastian.passehr@desy.de, schappacher@kabelbw.de}

\begin{abstract}
We present Version 9 of the Feynman-diagram calculator FormCalc and a 
flexible new suite of shell scripts and Mathematica packages based on 
FormCalc, which can be adapted and used as a template for calculations.
\hfill {\scriptsize Report DESY 2016-060, MPP-2016-63}
\end{abstract}

\section{Introduction}

The Mathematica package FormCalc \cite{FormCalc} simplifies Feynman 
diagrams generated by FeynArts \cite{FeynArts} up to one-loop order. 
It provides the analytical results and can generate Fortran or C code 
for the numerical evaluation of the squared matrix element.

This note presents new features in FormCalc 9:
\begin{itemize}
\item Combination of processes with (almost) arbitrary kinematics.

\item A wider range of cuts.

\item Improved code generation and driver programs.

\item Optional output of non-vectorized code.

\item Various QCD functions (running, $\MSbar\to\DRbar$, etc.).

\item Improvements in the makefiles and the build system.
\end{itemize}
Most of these improvements are the `collected wisdom' from several 
nontrivial SUSY calculations \cite{cs1,cs2,cs3,cs4} and provide, as it 
were, `extended support' for the MSSMCT model file \cite{MSSMCT}.

Furthermore a new suite of shell scripts and Mathematica packages based 
on FormCalc is introduced.  Presently used for the computation of the 
two-loop $\mathcal{O}(\alpha_t^2)$ Higgs-mass corrections in the MSSM, 
this code can easily be adapted and employed as a template for further 
calculations, and in this sense complements and generalizes what 
FormCalc does at one-loop.

\section{New Features and Improvements in FormCalc 9}

\subsection{Combining Processes}

FormCalc 9 allows to combine processes with (almost) arbitrary 
kinematics -- only the phase-space dimension must match, \ie the number 
of external legs must be the same.  This is mostly used in hadronic 
computations, but one could do \eg $e^+ e^-\to \{e^+ e^-, \mu^+\mu^-, 
\tau^+\tau^-\}$ in one go this way.

In a first step, the user generates code for all partonic processes 
needed with \Code{WriteSquaredME} as usual, though with output to 
separate folders via the \Code{Folder} option.  The symbol 
\Code{ProcName} may be used in the latter, as in:
\begin{alltt}
   WriteSquaredME[..., Folder -> \Brac{"squaredme", ProcName}]
\end{alltt}
and is substituted by the process name, abridged in such a way that it 
can safely be used as symbol or file name.

The file \Code{partonic.h} determines which processes are combined and 
how.  For each subprocess an entry of the following form must be added:
\begin{alltt}
#define PID 1
#define PARTON1 \Vi
#define PARTON2 \Vj
#include "\Var{folder}/specs.h"
#include "parton.h"
\end{alltt}
\begin{itemize}
\item \Code{PID} is an identifier enumerating the partonic processes 
  and runs from 1 to the number of processes (though not necessarily 
  in ascending order),
\item \Code{PARTON1} and \Code{PARTON2} give the PDG codes for the
  incoming partons, if the \Code{lumi\uscore hadron.F} module is used
  (they are ignored otherwise),
\item \Var{folder} is the folder into which code for the subprocess has 
  been written,
\item \Code{parton.h} comes with FormCalc (same for all subprocesses)
  and does the processing of the partonic information given.
\end{itemize}
Optionally, the user can choose in \Code{process.h} whether the results 
shall be combined at the differential or integrated level:
\begin{itemize}
\item \Code{JOIN\uscore PARTONIC} = 0, add differential partonic 
  cross-sections, then integrate sum (default),
\item \Code{JOIN\uscore PARTONIC} = 1, integrate each differential 
  partonic cross-section, then add up.
\end{itemize}
Once \Code{partonic.h} is complete it takes just the usual 
\Code{./configure} and \Code{make} to build the code.

\subsection{Veto Cuts}

Before FormCalc 9, only few one-particle cuts were available.  The 
selection was restricted since the cuts were applied by actually 
restricting the integration bounds, hence solvability of the cut 
equations was a limiting factor.

While these direct cuts are still available (and likely more efficient, 
in particular if large parts of phase-space are removed), a much wider 
range of so-called `veto cuts' has now been added.  They operate through 
an indicator (`veto') function which sets the cross-section to zero 
whenever a phase-space point removed by the cuts is sampled. The 
following one- and two-particle cuts are currently available and it is 
straightforward to add more in FormCalc's \Code{cuts.F}:

\begin{minipage}[t]{.3\hsize}
\begin{alltt}
CUT_E(\Vi)
CUT_k(\Vi)
CUT_ET(\Vi)
CUT_kT(\Vi)
CUT_y(\Vi)
CUT_eta(\Vi)
CUT_deltatheta(\Vi)
CUT_cosdeltatheta(\Vi)
\end{alltt}
\end{minipage}\begin{minipage}[t]{.5\hsize}
\begin{alltt}
CUT_R(\Vi,\Vj)
CUT_rho(\Vi,\Vj)
CUT_deltay(\Vi,\Vj)
CUT_deltaeta(\Vi,\Vj)
CUT_yprod(\Vi,\Vj)
CUT_etaprod(\Vi,\Vj)
CUT_deltaalpha(\Vi,\Vj)
CUT_cosdeltaalpha(\Vi,\Vj)
CUT_invmass(\Vi,\Vj)
\end{alltt}%
\end{minipage}

\medskip

\noindent
The overall cut condition is assembled from the one- and two-particle 
cuts in \Code{run.F}, in the definitions for \Code{CUT1...20}, for 
example
\begin{alltt}
#define CUT1 CUT_kT(3)\,.gt.\,10
\end{alltt}
requires the transverse momentum of particle \#3 to be greater than 10 
GeV.  The \Code{CUT1...20} definitions together make up one single 
logical expression and may include logical operators, \eg
\begin{alltt}
#define CUT1 CUT_kT(3) .gt. 10
#define CUT2 .and. CUT_kT(4) .gt. 10
\end{alltt}


\subsection{Non-vectorized Code}

FormCalc's \Code{WriteSquaredME} function generates code which is by 
default vectorized for the helicities of the external particles 
\cite{acat13}.

While vectorization is meanwhile almost universally available on all 
modern hardware, having this feature turned on for all code is not 
ideal, either.  For example, if the user wishes to include the generated 
subroutines in own code, the vectorization constitutes an extra layer of 
complexity.

For this reason FormCalc 9 allows to remove all vector instructions and 
special macros by choosing the output language with the \Code{"novec"} 
qualifier, viz.
\begin{alltt}
   SetLanguage["Fortran",\,"novec"]
   SetLanguage["C",\,"novec"]
\end{alltt}


\subsection{Debugging and Checking Instructions}

FormCalc has for long allowed to add debugging statements to the 
generated code, as in:
\begin{alltt}
   \Var{var} = \Var{expr}
   DEB("\Var{var}\/",\,\Var{var})
\end{alltt}
with (e.g.)
\begin{alltt}
#define DEB(\Var{tag},\Var{var}) print *, \Var{tag},\,"\,=\,",\,\Var{var}
\end{alltt}
This has been extended in Version 9 to include \emph{checking 
statements} of the form
\begin{alltt}
   CHK_PRE(\Var{var})
   \Var{var} = \Var{expr}
   CHK_POST("\Var{var}\/",\,\Var{var})
\end{alltt}
which might be implemented as
\begin{alltt}
#define CHK_PRE(\Var{var}) tmp = \Var{var}
#define CHK_POST(\Var{tag},\Var{var}) if(\,abs(\Var{var}\;-\,tmp)\,.gt.\,1D7\,) stop \Var{tag}
\end{alltt}
The idea here is to catch enormous variations from one parameter point 
to the next, which often point to some problem of the calculation, \eg 
an untreated resonance.  Finding the error locus is much more difficult 
using regular debug statements due to their copious output, which is
moreover difficult to compare using standard `diff' tools because even
results that are close (but not identical) numerically are usually 
highlighted in the diff output.

The user can individually choose the actual debugging statement 
(\Code{\$DebugCmd[\Var{i}\/]}) as well as commands issued before 
(\Code{\$DebugPre[\Var{i}\/]}) and after (\Code{\$DebugPost[\Var{i}\/]}) 
it, for \Var{i} = $1$ (debug), $2$ (pre-check), and $-2$ (post-check).


\subsection{New/improved QCD Functions}

Various QCD utility functions have been adapted, mostly from RunDec 
\cite{rundec}, and added to the `util' library of FormCalc.
They are used \eg for translating between various quantities in 
the MSSMCT model file \cite{MSSMCT}.
\begin{itemize}
\item subroutine \Code{AlphaS}:
  $\alpha_s^{\MSbar}(Q)$ up to 4-loop order \cite{rundec},

\item subroutine \Code{MqRun}:
  $m_q^{\MSbar}(Q_1)\to m_q^{\MSbar}(Q_2)$ up to 3-loop order \cite{rundec},

\item subroutine \Code{MqMSbar2OS}:
  $m_q^{\MSbar}(Q)\to m_q^{\text{OS}}$ up to 3-loop order \cite{rundec},

\item subroutine \Code{AsMSbar2DRbar}:
  $\alpha_s^{\MSbar}\to\alpha_s^{\DRbar}$ up to 2-loop order \cite{drbar}.

\smallskip

\item subroutine \Code{SetQCDPara}:
  set inputs ($\alpha_s(M_Z)$, thresholds) for the above routines.
\end{itemize}


\subsection{More Renormalization Constants}

The calculation of renormalization constants (RCs) has been extended to 
functions of arbitrary head, \eg mass shifts.  Output is written to a 
separate file for each head, for example:
\begin{alltt}
   RenConst[.] := ...  \(\to\)  RenConst.F,\,.h     {\it(so far)}
   MassShift[.] := ... \(\to\)  MassShift.F,\,.h    {\it(new)}
\end{alltt}
This effectively creates subsets of RC-like objects, \ie results of 
a loop calculation but constant (external kinematics fixed).
This is useful, for example, when computing the shifts used in the 
iterative determination of the Yukawa-resummation parameter $\Delta b$ 
since not all RCs, only the mass shifts, have to be recomputed in each 
iteration.

Results for head $h$ can either be requested explicitly by adding $h$
to the argument list, \eg
\begin{alltt}
   CalcRenConst[\Var{expr},\,MassShift]
\end{alltt}
or computed together with the others in the usual way (no extra 
arguments) by adding $h$ to the \Code{\$RenConst} list of RC-heads, \eg
\begin{alltt}
   $RenConst = \Brac{RenConst, MassShift}
\end{alltt}
A model file could for example request that \Code{MassShift} be computed 
by adding it to \Code{\$RenConst}.


\section{Extensions}

FeynArts/FormCalc have been designed, as many other software packages -- 
and not just in high-energy physics, to do a `complete' job.  That is, 
all the steps from the generation of Feynman diagrams to the numerical 
computation of a cross-section are executed from a single control 
program (\eg a single Mathematica session) -- that at least is how the 
demo programs insinuate usage.  There is nothing inherently wrong with 
this `monolithic' approach, it is maybe just not obvious how to extend 
it beyond the limitations of the package(s) used.

Sometimes one needs to use other packages for specific tasks, or there 
are special requirements or adaptations necessary, for example
\begin{itemize}
\item Resummations (\eg the $hbb$ Yukawa coupling in the MSSM),
\item Approximations (\eg the gaugeless limit),
\item K-factors,
\item Nontrivial renormalization.
\end{itemize}
In the following we shall introduce a suite of scripts and small 
Mathematica programs \cite{FHatat} made for the computation of the 
two-loop $\mathcal{O}(\alpha_t^2)$ corrections to the Higgs masses in 
the MSSM \cite{atat}, with optimized output suitable for inclusion in 
FeynHiggs \cite{fh}.  While each step is specific to the corrections 
computed here the code can be used as a template due to its modular 
structure and adapted with little effort for similar calculations.


\subsection{Components of the Calculation}
\label{sec:components}

The MSSM Higgs masses receive leading two-loop (2L) corrections of 
$\mathcal{O}(\alpha_s\alpha_t)$ and $\mathcal{O}(\alpha_t^2)$.
We are performing a diagrammatic calculation of the latter in the full 
complex MSSM in the gaugeless limit at $p^2 = 0$.  The following 
ingredients are needed:
\begin{enumerate}
\item[\ding{192}] The unrenormalized 2L self-energies
  $\Sigma_{hh}^{(2)}$, $\Sigma_{hH}^{(2)}$, $\Sigma_{hA}^{(2)}$,
  $\Sigma_{HH}^{(2)}$, $\Sigma_{HA}^{(2)}$, $\Sigma_{AA}^{(2)}$,
  $\Sigma_{H^+H^-}^{(2)}$
in gaugeless approximation at $p^2 = 0$ at leading order in
$\mathcal{O}(\alpha_t^2)$.

\item[\ding{193}] The 1L diagrams with insertions of 1L counterterms.

\item[\ding{194}] The 2L counterterms for \ding{192}.

\item[\ding{195}] The 2L tadpoles $T_h^{(2)}$, $T_H^{(2)}$, $T_A^{(2)}$
  at $\mathcal{O}(\alpha_t^2)$ appearing in \ding{194}.
\end{enumerate}


\subsection{Template for Calculations}

Our present code is based on several Mathematica Notebooks used in the 
former calculation \cite{atat}.  From the aspect of software development 
these Notebooks had various shortcomings: duplicate code (\eg gaugeless 
limit implemented multiply), parallel instructions poured all over, even 
the requirement to run with a particular Mathematica version -- all 
signs that the use of different packages together with various special 
requirements was nontrivial and that the reorganization was helpful.

We broke the calculation into several steps and implemented each step as 
an independent (stand-alone) shell script.  These scripts are run from 
the command line during development, while in the production run a 
makefile organizes the entire sequence.  In lieu of \textsl{in vivo} 
debugging, \eg setting breakpoints, which is not easily possible with a 
shell script we have set up detailed log files.  The outcome is a 
template for 2L calculations with optimized output in a nontrivial model 
with nontrivial renormalization.

The scripts use several external packages: FeynArts for the diagram 
generation \cite{FeynArts} using the MSSM model file with 1L 
counterterms MSSMCT.mod \cite{MSSMCT}, FormCalc for the 1L tensor 
reduction and code generation \cite{FormCalc}, and TwoCalc for the 2L 
tensor reduction \cite{TwoCalc}.


\newcommand\xbox[4]{%
  \hbox{\lower 6bp\hbox{\begin{picture}(50,18)
  \CBox(0,0)(50,18){#1}{#2}
  \Text(25,8)[]{\Black{\texttt{#3\vphantom{p}}}}
  \put(-5,4){\parbox[t]{60bp}{\begin{center}
    {\tiny\sf #4}
    \end{center}}}
  \end{picture}}}}

\newcommand\Xbox[4]{%
  \hbox{\lower 6bp\hbox{\begin{picture}(50,18)
  \CBox(0,0)(50,18){#1}{#2}
  \Text(25,8)[]{\Black{\texttt{#3\vphantom{p}}}}
  \put(-5,14){\parbox[b]{60bp}{\begin{center}
    {\tiny\sf #4}
    \end{center}}}
  \end{picture}}}}

\newcommand\cbox[2]{\xbox{Blue}{PastelBlue}{#1}{#2}}

\newcommand\Cbox[2]{\Xbox{Blue}{PastelBlue}{#1}{#2}}

\newcommand\gbox[2]{\xbox{White}{Gray}{#1}{#2}}

\newcommand\xrout[1]{%
  \begin{picture}(0,0)
  \Text(7,3)[l]{\tiny\sf #1}
  \end{picture}}

\newcommand\outtext[1]{{\scriptsize\sf #1}}

\newcommand\lout[1]{%
  \begin{picture}(0,0)
  \Text(-5,3)[r]{\outtext{#1 $\Rightarrow$}}
  \end{picture}}

\newcommand\rout[1]{%
  \begin{picture}(0,0)
  \Text(5,3)[l]{\outtext{$\Leftarrow$ #1}}
  \end{picture}}

\newcommand\trout[2]{%
  \begin{picture}(0,0)
  \Text(5,8)[l]{\outtext{$\Leftarrow$ #1}}
  \Text(5,0)[l]{\outtext{$\Leftarrow$ #2}}
  \end{picture}}

\bigskip
\bigskip

\begin{center}
\begin{tabular}{ccccc}
\lout{FeynArts}%
\Cbox{1-amps}{diagram generation} &
$\to$ &
\Cbox{2-prep}{preparation for\\[-2ex] tensor reduction} &
$\to$ &
\Cbox{3-calc}{tensor reduction}%
\trout{TwoCalc}{FormCalc} \\[1.5ex]
$\uparrow$ & & & & $\downarrow$ \\[.5ex]
\gbox{0-glmod}{model file preparation}%
\rout{MSSMCT.mod}
& & & & \cbox{4-simp}{}\xrout{simplification} \\[1.5ex]
& & & & $\downarrow$ \\[.5ex]
\lout{FormCalc}%
\cbox{7-code}{code generation} &
$\leftarrow$ &
\cbox{6-comb}{combination of results} &
$\leftarrow$ &
\cbox{5-rc}{calculation of\\[-2ex] renorm. constants}%
\rout{FormCalc}
\end{tabular}
\end{center}

\bigskip


\subsection{Script Structure}

The shell scripts (\verb=/bin/sh=) have several traits in common:
\begin{itemize}
\item Each script is run from the command line with up to two arguments, 
\eg
\begin{alltt}
   scripts/1-amps \Var{arg\(\sb1\)} \Var{arg\(\sb2\)}
\end{alltt}
\begin{tabbing}
where \=\Var{arg$_1$} =
\= \Code{h0h0}, \Code{h0HH}, \Code{h0A0}, \Code{HHHH},
   \Code{HHA0}, \Code{A0A0}, \Code{HmHp}
   (self-energies), \\
\>\> \Code{h0}, \Code{HH}, \Code{A0}
   (tadpoles), \\[.5ex]
\> \Var{arg$_2$} =
\> \Code{0}~~\= for virtual 2L diagrams, \\
\>\> \Code{1}  \> for 1L diagrams with 1L counterterms.
\end{tabbing}
That is, each of the components listed in Sect.~\ref{sec:components} is 
computed and stored individually.  Arguments are checked upon entry and 
meaningless arguments are omitted (\eg no \Var{arg$_2$} after 
combination of virtual and counterterm diagrams).

\item Inputs and outputs are defined in the first few lines, \eg
\begin{verbatim}
   in=m/$1/2-prep.$2
   out=m/$1/3-calc.$2
\end{verbatim}

\item Symbolic output including log files are written to the `\Code{m}' 
subdirectory.  The log file names are the output file names with 
\Code{.log.gz} appended.

\item Fortran code is written to the `\Code{f}' subdirectory.

\item The scripts use the shell's `here documents' to run the 
Mathematica Kernel \cite{LL2006}.
\end{itemize}


\subsection{Script Tasks}

In the following we shall briefly describe the main tasks of each 
script.  The total running time for all scripts for all 
self-energies/tadpoles (\ie the complete calculation) is about 15--20 
min.

\paragraph{Step 0:} Gaugeless Limit

We have to work in the gaugeless limit and also set $m_b = 0$ in order 
that the $\mathcal{O}(\alpha_t^2)$ corrections form a supersymmetric and 
gauge-invariant subset.  The gaugeless limit consists of \ding{192} 
setting the gauge couplings $g, g' = 0$ and hence $M_W, M_Z = 0$ while
keeping \ding{193} the weak mixing angle and \ding{194} the quantities
$\delta M_W^2/M_W^2$ and $\delta M_Z^2/M_Z^2$ finite.

No.\ \ding{192} and \ding{193} are most efficiently accomplished 
by modifying the Feynman rules, carried out by the \Code{0-glmod} 
script:
\begin{itemize}
\item Load the \Code{MSSMCT.mod} model file.
\item Modify couplings and remove those that become zero.
\item Write out an \Code{MSSMCTgl.mod} model file.
\end{itemize}


\paragraph{Step 1: Diagram Generation}

The \Code{1-amps} script generates the 2L virtual (\Var{arg$_2$} = 0) 
and 1L-plus-counterterm (\Var{arg$_2$} = 1) diagrams using wrappers 
around FeynArts functions provided by the \Code{FASettings.m} package.  
The wrappers mainly simplify the diagram selection, \eg the following 
list of filters, one for each topology, is composed from elementary 
functions like \Code{htb[\Vi]} which restrict the particles on the 
\Vi-th propagator:

\begin{minipage}{.5\hsize}
\begin{picture}(0,0)
\SetOffset(135,-68)
\SetColor{Blue}
\Curve{(0,0)(18,10)(35,15)(109,11)(126,5)}
\COval(126,4)(4,4)(0){Blue}{White}
\COval(2,0)(10,20)(0){Blue}{White}
\Text(65,8)[b]{\tiny ``one of $h_i,\tilde\chi,$}
\Text(65,-1)[b]{\tiny $t,\tilde t,\ b,\tilde b$''}
\end{picture}%
\begin{verbatim}
   sel[0][S[_] -> S[_]] = {
     t[3] && htb[6],
     t[3] && tb[6],
     t[3] && tb[6],
     t[3] && t[4] && htb[5],
     t[3] && htb[5|6],
     t[3] && htb[5],
     t[3] && t[5],
     t[5] && ht[3|4],
     t[3|4|5] && ht[3|4|5] }
\end{verbatim}
\end{minipage}\begin{minipage}{.4\hsize}
\includegraphics[width=\hsize]{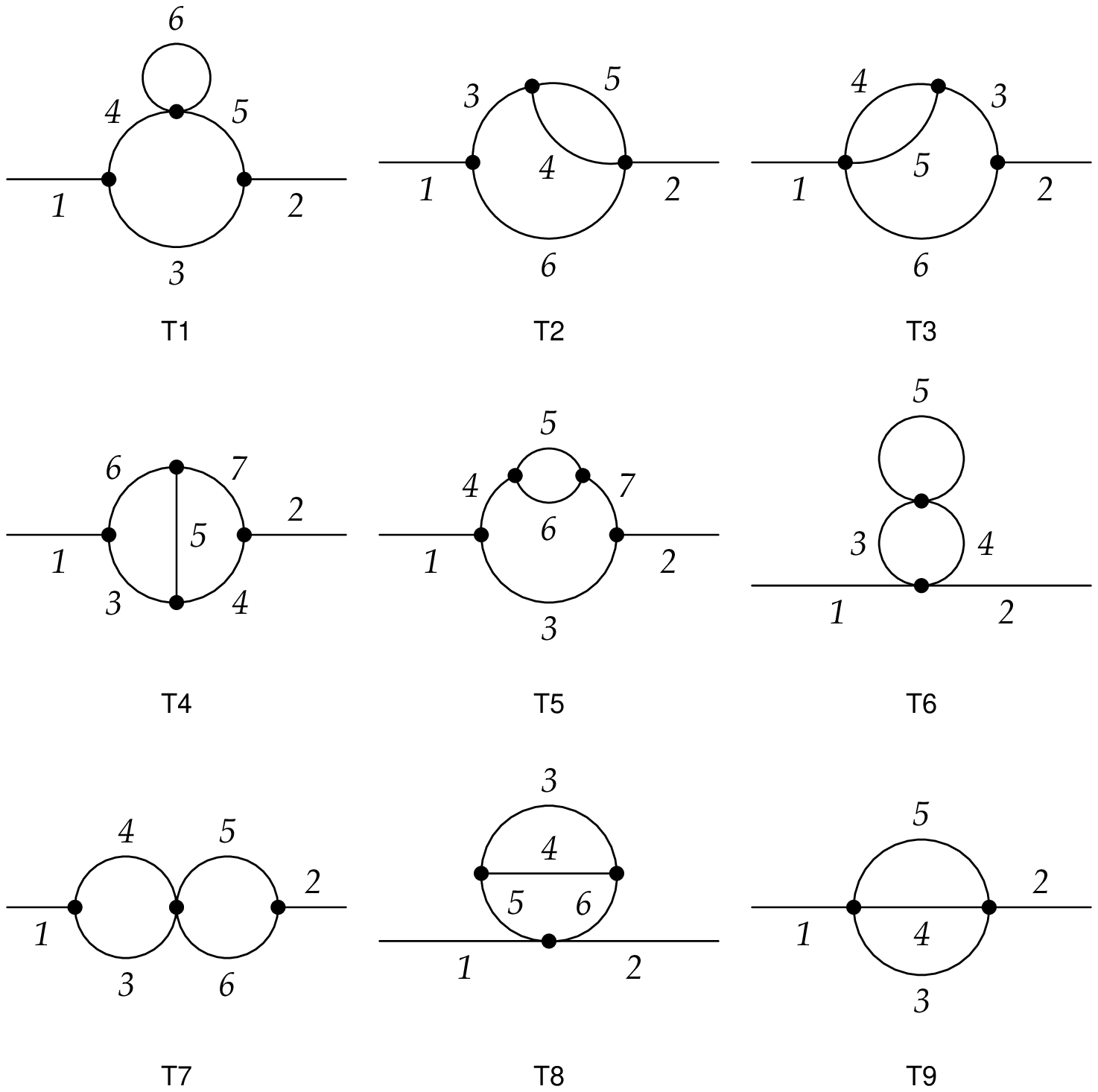}
\end{minipage}


\paragraph{Step 2: Preparation for Tensor Reduction}

The tensor reduction is traditionally the step which increases the 
number of terms in the amplitude most, therefore we inserted Step 2 
before and Step 4 after the tensor reduction to keep the number of terms 
as small as possible at all times.

Step 2, `before', takes the $p^2\to 0$ limit and simplifies the 
ubiquitous sfermion mixing matrices $U_{ij}$, mostly by exploiting 
unitarity ($\sim$ 50\% size reduction).

The unitarity relations of $2\times 2$ matrices are easily written down, 
\eg $U_{11} U_{11}^* + U_{12} U_{12}^* = 1$, but Mathematica rarely 
arranges expressions in just the way that they apply directly.  This can 
be improved by adding definitions for single summands, \eg $|U_{22}|^2 = 
|U_{11}|^2$.  Because they apply \emph{while} Mathematica ponders
the simplification strategy they increase the incentive for 
\Code{Simplify} to choose the unitarity-simplified version.  The product 
of two matrix elements is too `deep' for Mathematica to consider as an 
l.h.s., however, hence we must introduce intermediate symbols: we 
substitute $\Code{USf[1,\,\Var{j}\/]}~\Code{USfC[1,\,\Var{j}\/]} \to 
\Code{UCSf[1,\,\Var{j}\/]}$ etc.\ and can then formulate unitarity as 
\eg \Code{UCSf[2,2]} = \Code{UCSf[1,1]}.  The \Code{USfSimplify.m} 
package contains the complete set of such rules.


\paragraph{Step 3: Tensor Reduction}

The actual tensor reduction is a relatively straightforward invocation 
of TwoCalc \cite{TwoCalc} for the 2L and FormCalc \cite{FormCalc} for 
the 1L amplitudes.  Noteworthy here is perhaps that TwoCalc and FormCalc 
cannot be loaded into one Mathematica session because of symbol 
conflicts; this is easily accomodated in the shell-script setup, 
however.


\paragraph{Step 4: Simplification}

The amplitudes are simplified again after tensor reduction using a 
largely empirical recipe.  The following trick, lifted from 
\cite{diagmark}, enhances simplification efficiency greatly: already 
during diagram generation a function of the form 
\Code{DiagMark[\Var{m$_i$}]} has been inserted by the FeynArts wrapper 
functions, where $m_i$ are the masses running in the loop.  Few 
simplifications can be made between parts with different 
\Code{DiagMark}, hence simplification can be restricted to the 
prefactors of the \Code{DiagMark}, rather than the entire expression, 
which is of course much faster.


\paragraph{Step 5: Calculation of Renormalization Constants}

The 1L RCs are computed with FormCalc.  As mentioned in Step~0, $\delta 
M_V^2/M_V^2$ ($V = W, Z$) must remain finite in the gaugeless limit in 
which $M_V$ vanishes.  By substituting the explicit mass dependence: 
$\Code{dMVsq1}\to \Code{MV2~dMVsq1MV2}$ the masses cancel and we can 
take the limit safely.

Furthermore we insert expansions of the loop integrals in $\varepsilon = 
\frac{4 - D}2$ (package \Code{ExpandDel.m}) and collect the RCs w.r.t.\ 
powers of $\varepsilon$.  For example, the 1L top-mass counterterm 
becomes
\begin{alltt}
   \Brac{ dMf1[3,3] \(\to\) RC[-1, dMf1[-1,3,3]] + RC[0, dMf1[0,3,3]],
     \Brac{dMf1[-1,3,3] \(\to\) ..., dMf1[0,3,3] \(\to\) ...} }
\end{alltt}
where the first line denotes the expansion in $\varepsilon$ (the interim 
function \Code{RC} is used for power counting later) while the second 
gives the actual expressions for the coefficients.  This enables us to 
write down just the $\varepsilon^0$-coefficient of the final result in 
Step 6 without substituting back all RCs.


\paragraph{Step 6: Combination of Results}

The virtual 2L and 1L-plus-counterterm amplitudes, computed separately 
so far, are combined with the genuine 2L counterterms (hand-coded, in 
\Code{MSSMCT.rc2}) into one expression for each self-energy/tadpole.  We 
insert the $\varepsilon$-expansions of the RCs from Step 5 and once more 
those of the loop integrals, and extract the coefficient of 
$\varepsilon^0$ (or of $\varepsilon^{-1}$ or $\varepsilon^{-2}$ if a 
debug flag is set).  This expression is simplified again and forms then 
the final analytical result.


\paragraph{Step 7: Code Generation}

The renormalized self-energies are grouped into three subroutines for 
output, \Code{TLsp\uscore atat\uscore c} for the ones needed for 
renormalization (\Code{A0A0}, \Code{HmHp}), \Code{TLsp\uscore 
atat\uscore e} for the CP-even ones, and \Code{TLsp\uscore atat\uscore 
o} for the CP-odd ones.  These are the units in which FeynHiggs needs to 
compute the self-energies based on its flags.

Abbreviations are introduced for loop integrals and common 
subexpressions to shorten the overall expression ($\sim$ 50\% size 
reduction).  Optimized Fortran code is then written out using FormCalc's 
\Code{WriteExpr} and ancillary functions \cite{acat2010}.  Finally, 
static code is added which embeds the generated routines into FeynHiggs.  
The Fortran output directory `\Code{f}' totals about 350 kBytes in size 
and can directly be moved into the FeynHiggs source tree.


\section{R\'esum\'e}

FormCalc Version 9, available for download from 
\Code{http://feynarts.de/formcalc}, introduces various new features, 
most adapted/generalized from real-life SUSY calculations:
\begin{itemize}
\item Combination of processes,
\item More cuts,
\item Improved code generation,
\item New/improved driver and utility programs.
\end{itemize}
As a showcase of how to flexibly use FeynArts and FormCalc together with 
other packages, a suite of shell scripts and Mathematica packages was 
presented, originally made for the computation of the two-loop 
$\mathcal{O}(\alpha_t^2)$ Higgs-mass corrections, which may serve as 
template for similar calculations:
\begin{itemize}
\item Two-loop,
\item Nontrivial model (MSSM) and renormalization,
\item Specific approximations (gaugeless, $\smash{p^2} = 0$),
\item Optimized, compact output.
\end{itemize}
The code is included in FeynHiggs 2.11.0 and above in the 
\Code{gen/tlsp/} subdirectory.


\bigskip
\raggedright

\end{document}